\documentclass[]{spie}  

\usepackage{amsmath,amsfonts,amssymb}
\usepackage{graphicx}
\usepackage[colorlinks=true, allcolors=blue]{hyperref}
\usepackage{multicol}
\usepackage{subfigure}
\usepackage{wrapfig}
\usepackage[final]{microtype}

\usepackage[ruled,vlined,linesnumbered]{algorithm2e}

\SetCommentSty{mycommfont}

\usepackage{amsmath}

\title{DrDisco: \underline{D}eep \underline{R}egistration for \underline{Dis}tortion \underline{Co}rrection of Diffusion MRI with single phase-encoding}

\author[a]{Zhangxing Bian}
\author[a]{Muhan Shao}
\author[a]{Aaron Carass}
\author[a]{Jerry L. Prince}
\affil[a]{Department of Electrical and Computer Engineering, Johns~Hopkins~University,~Baltimore,~MD~21218,~USA}

\authorinfo{Further author information: (Send correspondence to Zhangxing Bian.)\\Zhangxing Bian: E-mail: zbian4@jhu.edu}

\begin{document} 
\maketitle

\begin{abstract}

Diffusion-weighted magnetic resonance imaging~(DW-MRI) is a non-invasive way of imaging white matter tracts in the human brain. DW-MRIs are usually acquired using echo-planar imaging~(EPI) with high gradient fields, which could introduce severe geometric distortions that interfere with further analyses. Most tools for correcting distortion require two minimally weighted DW-MRI images (B0) acquired with different phase-encoding directions, and they can take hours to process per subject. Since a great amount of diffusion data are only acquired with a single phase-encoding direction, the application of existing approaches is limited. We propose a deep learning-based registration approach to correct distortion using only the B0 acquired from a \textit{single} phase-encoding direction. Specifically, we register undistorted T1-weighted images and distorted B0 to remove the distortion through a deep learning model. We apply a differentiable mutual information loss during training to improve inter-modality alignment. Experiments on the Human Connectome Project dataset show the proposed method outperforms SyN and VoxelMorph on several metrics, and only takes a few seconds to process one subject.

\end{abstract}


\section{Introduction}

Diffusion-weighted magnetic resonance imaging~(DW-MRI) is able to image the diffusion of water molecules in biological tissues in a non-invasive way \cite{le1986mrShort,taylor1985spatial}. This allows us to infer information about tissue microstructure~\cite{novikov2019quantifying} and the anatomical connections of the brain in a technique known as fiber tractography~\cite{mori2002fiber, mori1999three, xue1999vivo, bian2023spieB}. 
DW-MRIs are usually acquired using echo-planar imaging~(EPI)\cite{ordidge1999development} with high gradient fields, which allow a large number of diffusion-weighted images to be acquired in a short period of time. Such acquisitions suffer from geometric distortions, and signal losses are introduced due to susceptibility-induced field defects, both of which hamper further image analyses~\cite{wu2008comparisonShort}.

Existing tools such as TOPUP~\cite{andersson2003correctShort}, require minimally weighted DW-MRI images~(B0) acquired with reverse phase-encoding parameters (also known as ``blip-up and blip-down'') to estimate the static susceptibility field-maps and distortions. However, these supplementary data are not always available due to scanner limitations or acquisition protocols. The recent method Synb0-DISCO~\cite{schilling2020distortionShort} relaxes these constraints by synthesizing the B0 image from T1-weighted~(T1w) images using a generative adversarial network and then apply TOPUP, which takes hours to process a subject. 

Registration-based algorithms~\cite{wu2008comparisonShort, smith2004advancesShort} are able to directly correct geometric distortion by non-rigidly aligning distorted DW-MRIs and undistorted T1w images, thus circumventing the need to acquire reverse phase-encoding scans or field-maps. However, they suffer from inter-modality inaccuracies and slow speed. The alignment is problematic in areas with limited tissue contrast, and the widely-used SyN method~\cite{ avants2011reproducibleShort} can take tens of minutes. 

A recent learning-based framework, VoxelMorph~\cite{balakrishnan2019voxelmorphShort}, formulates registration as a function that transforms a pair of input images into a deformation field that aligns them. It can be trained on pairs of images to optimize typical image matching objectives under a smoothness constraint. VoxelMorph can quickly compute a deformation field given a new pair of scans by simply evaluating the function. Although the VoxelMorph framework has been widely used for intra-modality (i.e., same modality) registration, it is less optimal for inter-modality registration, which is required for the distortion correction task. 

Inspired by the aforementioned works, we propose a learning-based registration approach to correct the distortion of DW-MRI without needing reverse phase-encoding scans or field-maps. We incorporate a differentiable mutual information~(MI) loss during training to improve the inter-modality registration between T1w and B0 images. Our experiments on the Human Connectome Project~(HCP)~\cite{van2013wuShort} data show that our approach outperforms VoxelMorph and SyN on two metrics, while taking less than a minute of CPU time and less than two seconds of GPU time, while TOPUP-dependent approaches can take hours.

\begin{figure*}[!tb]
	\centering
		\includegraphics[width=\linewidth]{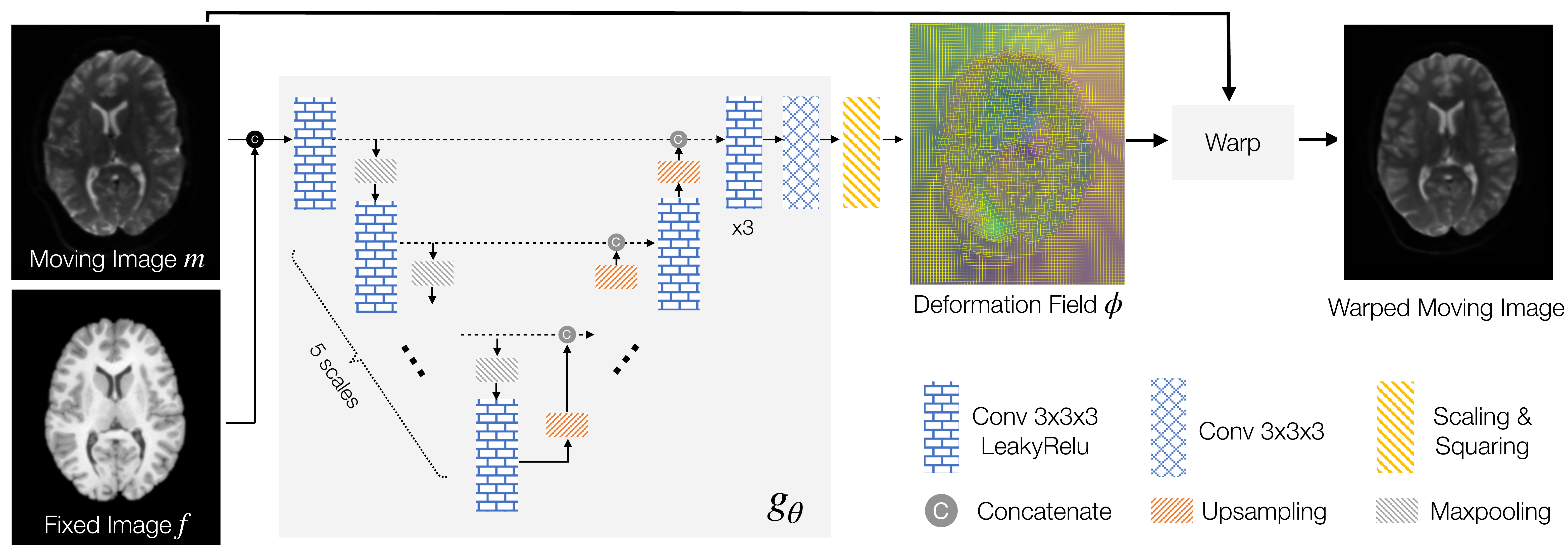}
		\caption
		{ The framework of our method. Given a fixed image, which is distortion-free, and a B0 image from the diffusion MRI, which suffers from geometric distortion, the Unet-like network predicts the deformation that corrects the distortion. 
		}  %
		\label{fig:Methods}
\end{figure*}

\section{Method}
\figurename~\ref{fig:Methods} shows the framework of our method. Formally, we denote $f$ and $m$ as two image volumes over a 3-D spatial domain $\Omega \subset \mathbb{R}^{3}$. The function $g_{\theta}(f,m) = \mathbf{\phi}$ gives the transformation $\phi$ that is used to align the fixed and moving images. The function $g$ is modelled by a convolutional neural network~(CNN) parametrized by learnable parameters $\theta$. The parameters are learned through optimizing a general objective: 
\begin{equation}
	\hat{\theta} =  \arg \min_{\theta} \mathcal{L}_{sim}(f, m \circ g_{\theta}(f,m)) + \lambda \mathcal{L}_{smooth}(g_{\theta}(f,m))\,.
\end{equation}
The first term, $\mathcal{L}_{sim}$, encourages image similarity between the fixed and warped moving image. The second term, $\mathcal{L}_{smooth}$, imposes a smoothness constraint on the transformation field. The parameter $\lambda$ determines the trade-off between these two terms. 
Intensity mean squared error~(MSE), MI~\cite{viola1997alignmentShort}, and cross-correlation~(CC)~\cite{avants2008symmetricShort} are three of the most commonly used metrics for $\mathcal{L}_{sim}$. MSE is generally good for intra-modality registration while the latter two are good for inter-modality registration. MI is an information-based similarity measure that is especially good for intermodality registration~\cite{avants2011reproducibleShort}, and this is what we focus on here. 

The MI between two images $A$ and $B$ is defined as:
\begin{equation}
	I(A, B)=\sum_{a, b} p(a, b) \log \frac{p(a, b)}{p(a) p(b)}.
	\label{eq:t-MI}
\end{equation}
The probability $p(a)$ is the fraction of voxels in image $A$ with intensity $a$ and $p(b)$ is analogously defined for image $B$.  $p(a, b)$ is the joint distribution of the intensities of the two images, $A$ and $B$. To compute Eq.~\eqref{eq:t-MI} directly, one must bin (quantize) the image intensities. Binning, however, is not a differentiable operation, which implies that Eq.~\eqref{eq:t-MI} cannot be used for training $g_\theta$ with backpropagation. 

To solve this problem, Parzen windowing is used to calculate the \textit{continuous} contribution of each voxel to a range of histogram bins instead of contributing only to the bin it falls into~\cite{thevenaz2000optimization, guo2019multi}. Accordingly, the intensity distribution $P_A(x)$  of a single image $A$ at intensity value $x$ is:
	\begin{equation}
			P_A(x)=\frac{1}{n} \sum_{s \in A} W(x-s)\,,
		\end{equation}
Each sample's contribution is weighted by a Gaussian function, $W$, of its distance to $x$:
	\begin{equation}
		W(x-s)=\frac{1}{\sigma \sqrt{2 \pi}} e^{-\frac{(x-s)^{2}}{2 \sigma^{2}}}.
		\label{eq:Pwin}
	\end{equation}
%
 The intensity distribution $P_B(x)$ for image $B$ is obtained similarly. The joint probability is then calculated as:
\begin{equation}
	P_{A, B}(x, y)=\frac{1}{n} \sum_{(a, b) \in(A, B)} W(x-a) W(y-b).
\end{equation}

By plugging $P_A(x)$, $P_B(y)$, and $P_{A,B}(x,y)$ into Eq.~\eqref{eq:t-MI}, we obtain the differentiable mutual information~(dMI) loss $\mathcal{L}_{\operatorname{dMI}}(A, B) = -\tilde{I}(A, B)$. The negative sign is added for minimization purposes. We use the same smoothness loss as in VoxelMorph. Trilinear interpolation is used in warping. Our total loss is $\mathcal{L} = \mathcal{L}_{\operatorname{dMI}} + \lambda\mathcal{L}_{\operatorname{smooth}}$. 

\section{Experiments}

\textbf{Dataset}. We selected 440 subjects who have DW-MRI acquisition sessions from the HCP~\cite{van2013wuShort} dataset for training and evaluation. T1w images were preprocessed following the official HCP pipeline~\cite{glasser2013minimalShort}. B0 images~(the volume with b-value $< 50$) are extracted and averaged for each DW-MRI sequence. We carried out affine spatial normalization between B0 and T1w, and cropped the resulting images to 144$\times$176$\times$160. The dataset was split into 264, 88, 88 for training, validation, and testing, respectively. 

\begin{figure}[!tb]

	 		\centering
	 		\includegraphics[width=0.8\textwidth]{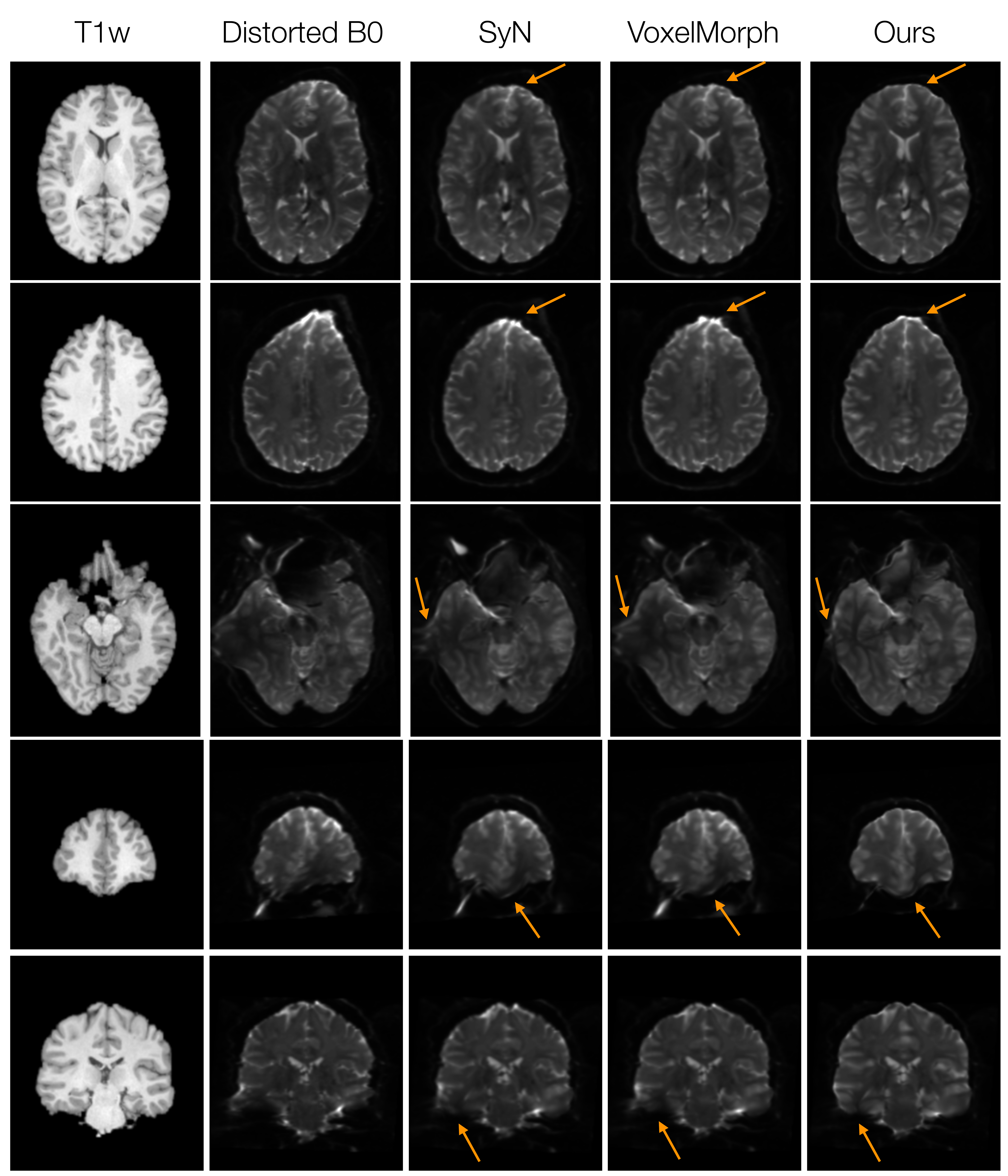}
	 		\caption{Qualitative comparison. The first two columns are the T1w (fixed) and B0 (moving) images. The next three columns show the correction results by SyN, VoxelMorph, and our method~(Ours).}%
	 		\label{fig:T1-b0}

	 	\end{figure}

\noindent\textbf{Evaluation}. We evaluate our model and report results on the test set and compare our method to SyN and VoxelMorph. As shown in Fig.~\ref{fig:T1-b0}, our approach outperforms SyN and VoxelMorph in challenging cases with large distortions, especially in the most distorted regions---the frontal lobe and ventral temporal lobe.

\begin{figure*}[!tb]
	\centering
	{\includegraphics[width=0.30\textwidth]{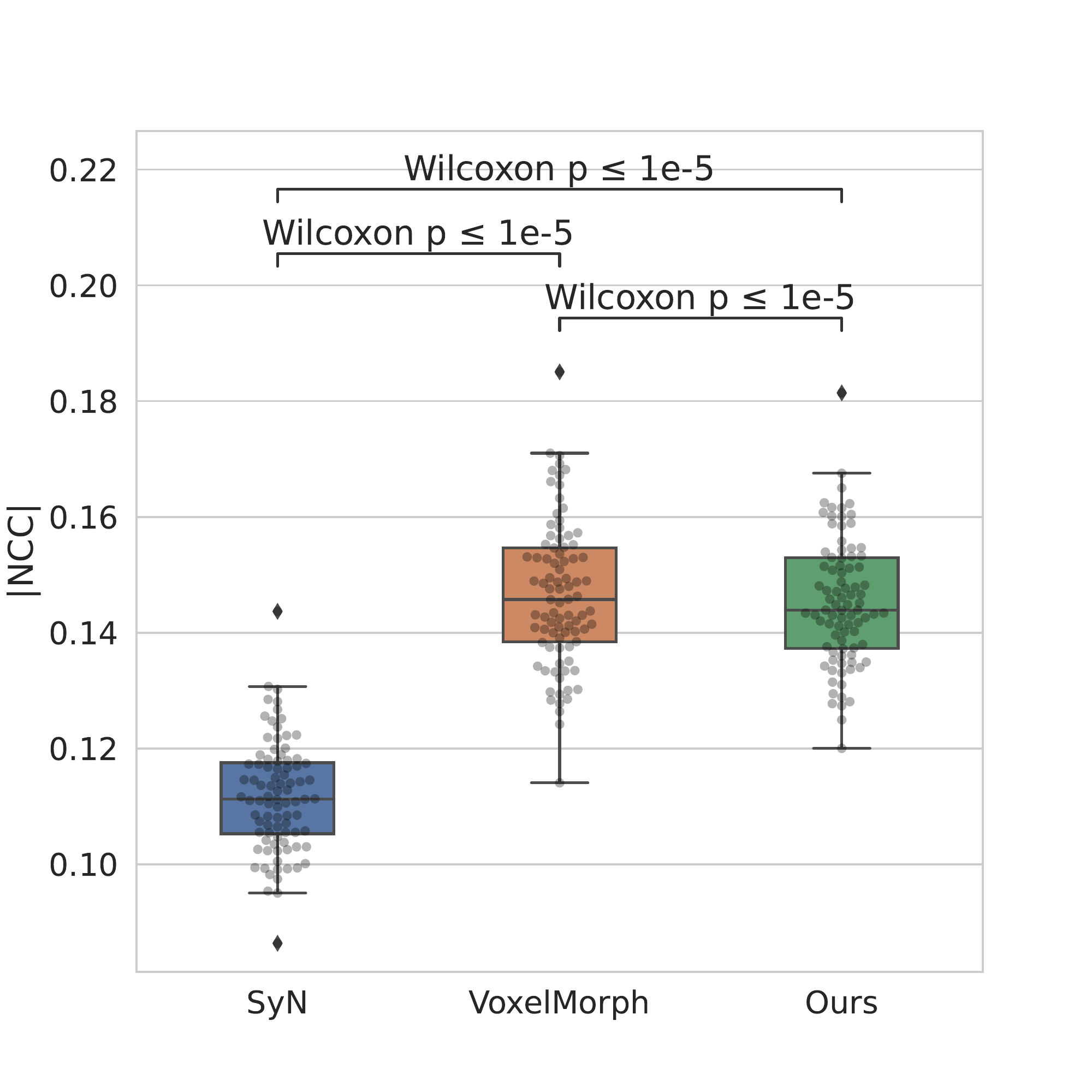}} 
	{\includegraphics[width=0.30\textwidth]{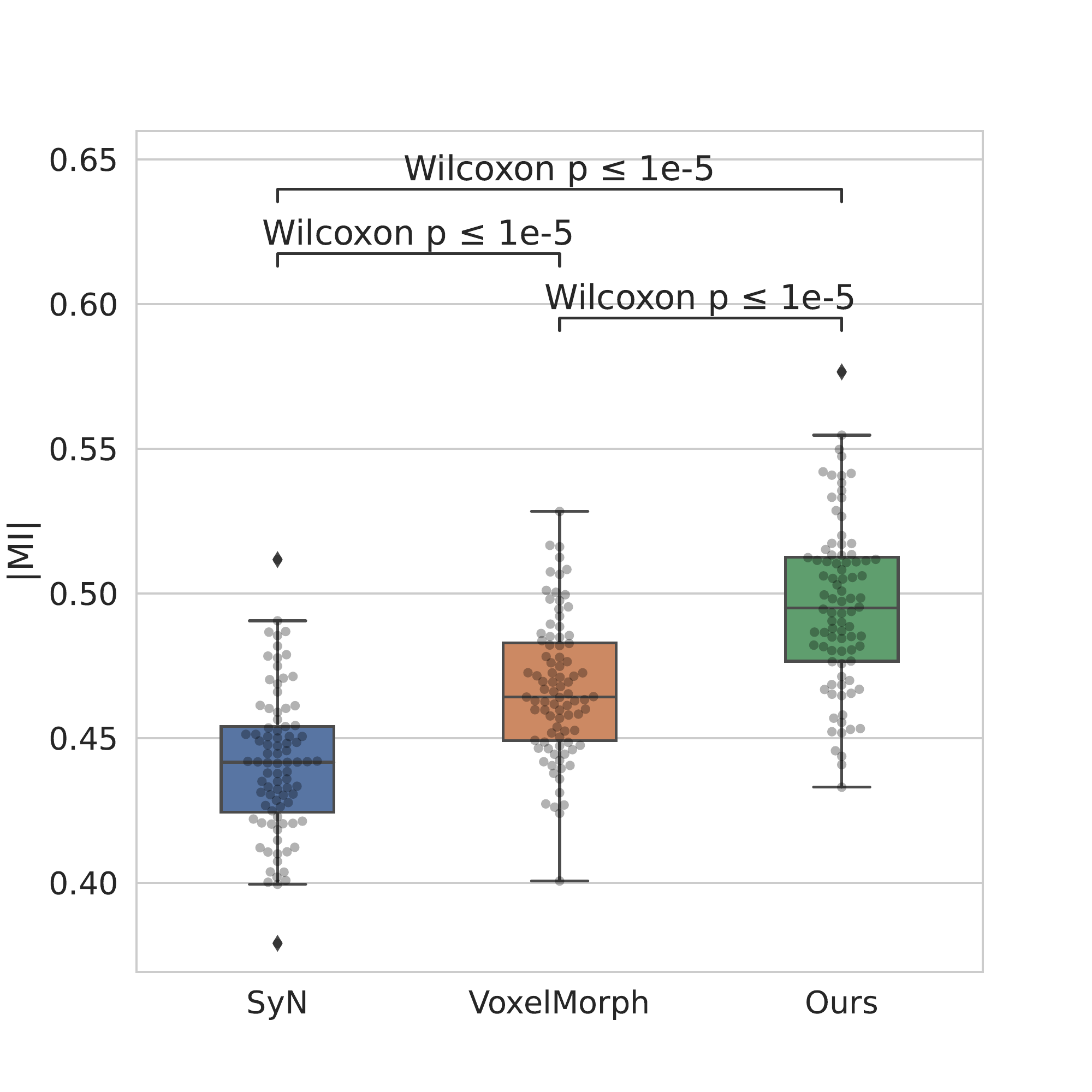}} 
	{\includegraphics[width=0.30\textwidth]{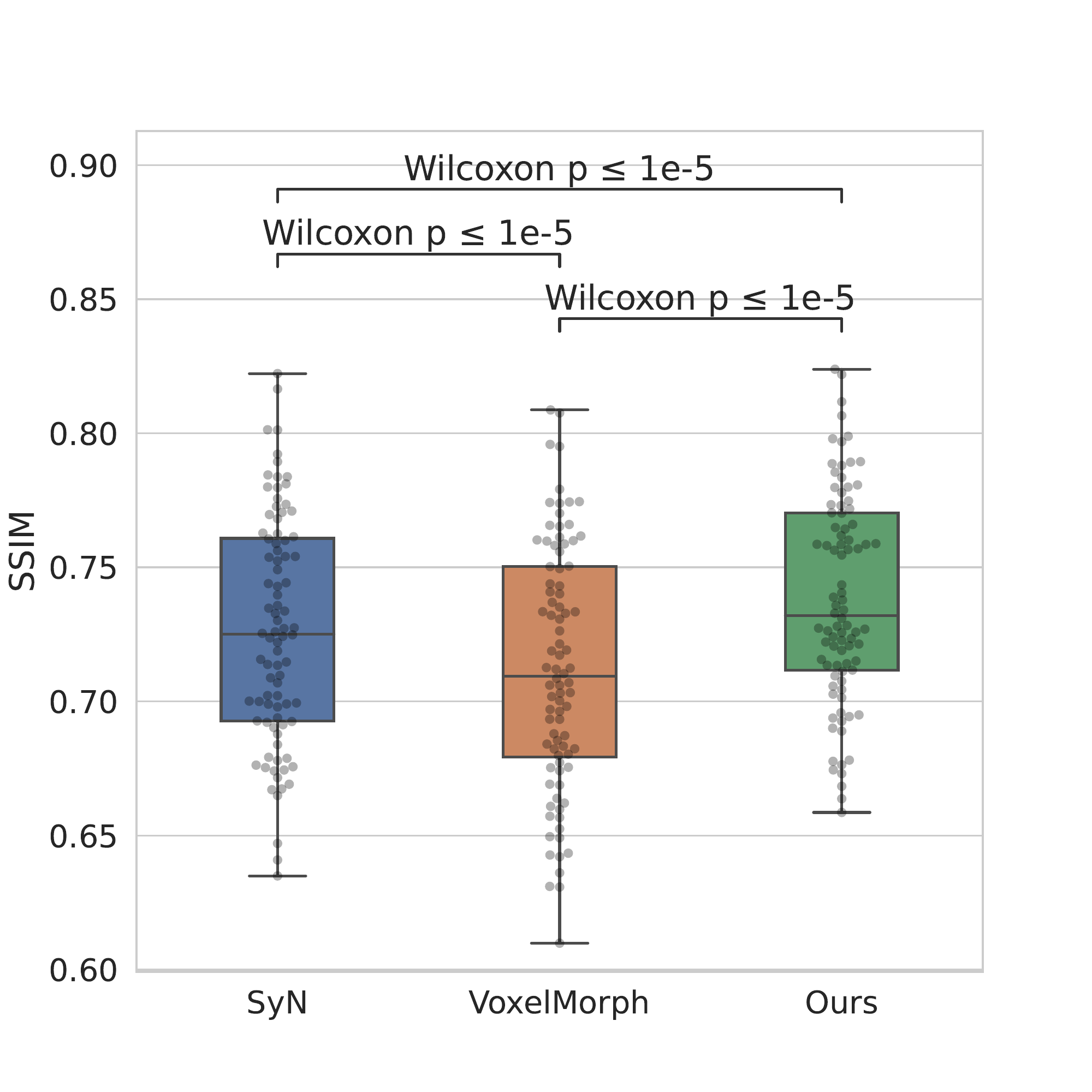}} 
	\caption{Quantitative comparison. We measure the similarity between distorted B0 and warped T1w through three metrics: SSIM~$\uparrow$, $\lvert \mathrm{NCC} \rvert$~$\uparrow$, and $\lvert \mathrm{MI} \rvert$~$\uparrow$. 88 different subjects from test set are evaluated. $\uparrow$ indicates the higher is better.}%
	\label{fig:boxplot}
\end{figure*}

It is impossible to obtain the ground truth of the deformation field since different registration fields can yield similar-looking warped images. So we measured the performance of distortion correction by measuring the similarity between the corrected B0 image and the T1w through three metrics: MI, Normalized Cross-correlation~(NCC), and structural similarity index measure~(SSIM). Note that the MI used here is not the dMI used for training our network, while the NCC is exactly the same as the NCC loss used for training VoxelMorph. \figurename~\ref{fig:boxplot} shows our method performs slightly worse than VoxelMorph on NCC, which is probably due to VoxelMorph directly optimizing NCC during training. However, our model performs significantly better than VoxelMorph and SyN on SSIM and MI. 

\noindent\textbf{Implementation Details}. We train both VoxelMorph\cite{balakrishnan2019voxelmorphShort} and the proposed method from scratch and determine the best $\lambda$ for each model through a hyper-parameter search. We choose the weight of smoothness as $\lambda = 0.7$ with NCC loss for VoxelMorph, and $\lambda = 0.3$ with the differentiable MI loss for the proposed method. We use a similar CNN backbone as VoxelMorph. Deep networks are trained until no further improvement is observed. Our implementation is in PyTorch. During testing, our method can process one case in less than 2s on a GPU~(2080~Ti).

\section{Discussion and Conclusion}  
We applied a differentiable MI loss in training a CNN model to correct the distortion for DW-MRIs without using field-maps or reverse phase-encoding scans. We outperform SyN and VoxelMorph on MI and SSIM. Our approach might be further improved by using newer deep learning registration techniques~\cite{liu2022mmmi}. Despite the preliminary success in correcting geometric distortion, the non-uniform magnetic field will also cause a loss of signal or signal pileup~(extreme bright spot), which cannot be recovered by registration only. We think correcting distortion and the signal loss and pileup simultaneously is a promising future direction.

\section*{Acknowledgements}
We want to thank Dr.~Blake Dewey (Johns Hopkins University) for insightful discussions.

\small
\bibliography{main} 
\bibliographystyle{spiebib} 

\end{document}